\documentclass[final,5p,times,twocolumn]{elsarticle}
\usepackage{amssymb,amsmath}

\begin{document}

\begin{frontmatter}

\title{On the inconsistency of Porsch-R\"oseler cutoff theory.
(reply to [1]).}

\author[lak]{V.D.~Lakhno\corref{cor1}}
\ead{lak@impb.psn.ru}

\cortext[cor1]{Corresponding author}

\address[lak]{Institute of Mathematical Problems of Biology, Russian Academy of Sciences, Pushchino, Moscow Region, 142290, Russia}

\begin{abstract}
It is shown that the conclusion made by Porsch and R\"oseler that their cutoff polaron theory transforms to Tulub theory is erroneous. The results of the work by Klimin and Devreese based on the results of Porsch and R\"oseler theory are erroneous too.
 
\end{abstract}

\begin{keyword}
cutoff parameter \sep bipolaron \sep Tulub approach
\end{keyword}

\end{frontmatter}

In paper \cite{1} a conclusion was made about "incompleteness" of translation-invariant continuum polaron theory developed by Tulub \cite{2}. The authors of \cite{1} drew this conclusion relying on paper \cite{3} where the results of Tulub theory are just reproduced. In this paper we will demonstrate that actually Porsch and R\"oseler theory containing a cutoff does not reproduce the results of \cite{2}.

As stated in \cite{1}, the functional of the total energy obtained by Tulub should contain an extra term $\delta E^{(PR)}_{R}$  which arises if an external cutoff is inserted in the continuum model. Such an external cutoff was considered in \cite{3}. For this purpose the total energy functional was integrated there with respect to phonon wave vectors not in the infinite limit but in the limit of a certain finite $q_0$ . As pointed out in \cite{3}, in the limit of $q_0\rightarrow\infty$  the quantity  $\delta E^{(PR)}_{R}\rightarrow 0$. On that ground the authors of \cite{3} decided that their theory is consistent with the results of theory \cite{2}.

To demonstrate that this is not the case we  will proceed from the expression for  $\delta E^{(PR)}_{R}$ \cite{3} which is basic in \cite{1} too:

\begin{eqnarray}
\label {eq.1}
\delta E^{(PR)}_{R}(q_0)=\frac{3\hbar}{2}(\Omega_{q_0}-\omega_{q_0}),
\end{eqnarray}

$$\Omega_{q_0}=\left\{\omega_{q_0}^2+\int _0^1 d\eta \int _0^{q_0}d q \,\frac{\hbar q^4f(q) \omega_q}{3\pi^2m}
\,\frac{2ReF(\omega_q+i\delta)+\left|F(\omega_q+i\delta)\right|^2}{\left|1+F(\omega_q+i\delta)\right|^2}\right\}^{1/2},$$

$$F(z)=\eta \frac{\hbar}{6\pi^2m}\int _0^{q_0}d q q^4 f^2 (q) \left(\frac{1}{\omega_q+z}+\frac{1}{\omega_q-z}\right),$$

$$f_q=-\left(V_q/\hbar \omega _0\right)exp \left(-q^2/2a^2\right),$$
where  $\omega_q=\omega_0+\hbar q^2/2m$,  $\omega_0$-is the LO - phonon frequency,  $V_q$-are the amplitudes of electron-phonon interaction, $a$- variational parameter. In the limit $\delta\rightarrow 0$ from (1) we have:

\begin{eqnarray}
\label {eq.2}
\Omega_{q_0}^2-\omega_{q_0}^2=\int _0^{q_0}d q \,\frac{\hbar q^4f(q)^2 \omega_q}{3\pi^2m}.
\end{eqnarray}
With the use of expression for $V_q$ : $V_q=(2\pi\sqrt{2})^{1/2}\alpha^{1/2}\sqrt[4]{\hbar^5\omega^{3}_{0}/m}\cdot1/q$, where $\alpha$ is a constant of electron-phonon coupling, from (1), (2) we get:

\begin{eqnarray}
\label {eq.3}
\delta E^{(PR)}_{R}=\frac{3\hbar}{2}\omega_{q_0}(\sqrt{1+\Delta/\omega^2_{q_0}}-1),
\end{eqnarray}

$$\Delta=\frac{2\sqrt{2}}{3\pi}\alpha\omega^2_0\left(\frac{\hbar a^2}{m\omega_0}\right)^{3/2}N_1\left(1+\hbar a^2N_2/2m\omega_0N_1\right),$$

$$N_1=\int^{q_0/a}_0dx\,x^2exp(-x^2), \quad N_2=\int^{q_0/a}_0dx\,x^4exp(-x^2).$$
Let us consider various limiting cases of expression (3). In the limit of small values of the cutoff parameter $q_0<<a$, from (3) we derive:

\begin{eqnarray}
\label {eq.4}
\Delta=\frac{2\sqrt{2}}{9\pi}\alpha\omega^2_0\left(\frac{\hbar q^2_0}{m\omega_0}\right)^{3/2}\left[1+\frac{3}{10}\left(\frac{\hbar q^2_0}{m\omega_0}\right)\right],
\end{eqnarray}
Formula (3), (4) imply that in this limit  $\delta E^{(PR)}_{R}$ is independent of parameter $a$. For $q_0>>a$, it follows from (3) that:

\begin{eqnarray}
\label {eq.5}
\delta E^{(PR)}_{R}=\frac{3\hbar}{2}\omega_{q_0}\left(\sqrt{1+q^4_{0c}/q^4_0}-1\right),
\end{eqnarray}

$$q_{0c}=\left(\alpha a^5/\sqrt{2\pi}\right)^{1/4}\left(\hbar/m\omega _0\right)^{1/8}$$
The value of $q_{0c}$  has the meaning of a quantity for which the integrand in Tulub's functional has a maximum \cite{2,4}. Notice that expression (8) presented in \cite{1} is derived from (5) only on condition that $a<<q_0<<q_{0c}$. From (5) it follows that in the limit $q_0>>q_{0c}$: 
\begin{eqnarray}
\label {eq.6}
\delta E^{(PR)}_{R}=\frac{3\hbar}{4}\omega_{q_0}{q^4_{0c}/q^4_0},
\end{eqnarray}
i.e. for large  $q_0$ the quantity $\delta E^{(PR)}_{R}\propto q^{-2}_{0}$. For this reason the authors of \cite{3} decided that theory \cite{3} transforms to theory \cite{2} with $\delta E^{(PR)}_{R}=0$. However, one can readily see that this is not the case. Indeed, in the functional of Tulub's total energy obtained by the authors of \cite{3}, the upper limit of integration must satisfy the condition  $q_0>q_{0c}$ which will automatically lead to the dependence of the polaron energy on the coupling constant $E\propto \alpha^{4/3}$ \cite{4}. Hence, Porsch-R\"oseler cutoff theory in no case reproduces Tulub theory. The use of such an inconsistent theory has lead the authors of \cite{1} to the conclusion of the "incompleteness" of theory \cite{2}. 

To sum up, let us formulate our final conclusions. As we have shown, the conclusion by Porsch and R\"oseler that $\delta E^{(PR)}_{R}$  vanishes as $q_0\rightarrow\infty$  is correct. This result, nevertheless, does not lead to Tulub theory. The statement that functional of Tulub's total energy is incomplete made in \cite{1} is erroneous. 

Calculations carried out by the author \cite{5}-\cite{7} on the basis of Tulub's approach \cite{2} do not give rise to doubt.

The work was supported by the Russian Foundation for Basic Research, Project nos. 11-07-12054 and 10-07-00112.

\end{document}